\documentstyle[prl,aps,twocolumn,floats,epsf]{revtex} 

\newcommand{\plb}[2]{{\em Phys. Lett.}              {\bf #1B}, #2 }
\newcommand{\npb}[2]{{\em Nucl. Phys.}              {\bf B#1}, #2 }

\newcommand{\prt}[2]{{\em Phys. Rev.}               {\bf D#1}, #2 }

\newcommand{\be}{\begin{equation}}
\newcommand{\ee}{\end{equation}}
\newcommand{\ba}{\begin{eqnarray}}
\newcommand{\ea}{\end{eqnarray}}

\def\al2{\frac{\alpha^2}{\pi^2}}

\begin{document}

\preprint{
\noindent
\hfill
\begin{minipage}[t]{3in}
\begin{flushright}
\vspace*{2cm}
\end{flushright}
\end{minipage}
}

\draft

\title{Two-loop Renormalization Group Equations in the Standard Model}
\author{Mingxing Luo and Yong Xiao}
\address{ Zhejiang Institute of Modern Physics, Department of Physics,
Zhejiang University, Hangzhou, Zhejiang 310027, P R China}

\date{July 2002}

\maketitle

\begin{abstract}
Two-loop renormalization group equations in the standard model 
are re-calculated.
A new coefficient is found in the $\beta$-function of the quartic coupling 
and a class of gauge invariants are found to be absent in the $\beta$-functions of 
hadronic Yukawa couplings. 
The two-loop $\beta$-function of the Higgs mass parameter is presented in complete form. 
\end{abstract}


Analysis based upon renormalization group equations (RGEs) plays an important role in the
study of physics of the standard model (SM) and beyond.
Detailed analysis of RGEs confirmed the behavior of asymptotic freedom in QCD,
and thus helped to
establish a non-Abelian gauge theory for the strong interaction \cite{pdg}.
The runnings of coupling constants and mass parameters are crucial in global analysis of 
high precision electroweak experiments \cite{erler}.
On the other hand, RGEs analysis extrapolated to extremely high energy provides a
possible test for physics beyond the SM. 
For example, gauge couplings do not unify within the SM.
This gives extra evidence against simple grand unification theories such as $SU(5)$ without supersymmetry,
in addition to the non-observation of proton decay.
On the other hand, gauge couplings seem to unify 
at a scale $\sim 2\times 10^{16} \ GeV$ in the minimal
supersymmetric standard model, which can be interpreted 
as an indirect evidence for supersymmetry as well as
unification theories \cite{GUT1,GUT2,GUT3}. Comprehensive analysis can be found in \cite{rammond}.

Computations of RGEs in gauge theories have been performed for various models 
to different orders of perturbation.
Persistent efforts yielded recently a four-loop result of the 
$\beta$-function of the strong coupling constant \cite{4loops}.
Two-loop RGEs of dimensionless couplings in a general gauge theory
as well as the specific case of the SM had been 
calculated long ago in a series
of classic papers by Machacek and Vaughn \cite{MV1,MV2,MV3}.
By introducing a non-propagating gauge-singlet ``dummy" scalar field, 
two-loop RGEs of dimensional couplings
can be readily inferred from dimensionless results \cite{martin,luo}.
These were used to derive the RGEs of supersymmetric theories 
a decade later \cite{martin}.

In this paper we re-calculate the two-loop RGEs in the SM, in a combination
of using the general results of \cite{MV1,MV2,MV3} and direct calculations from
Feynman diagrams.
A new coefficient is found in the $\beta$-function of the quartic coupling
and a class of gauge invariants are found to be absent in $\beta$-functions of
hadronic Yukawa couplings.
We will also present the two-loop $\beta$-function of the Higgs mass parameter 
in complete form, which provides a partial but useful check on
the calculation of the quartic coupling. 
Whenever discrepancy with the literature appears, we carefully inspect
relevant Feynman diagrams to ensure consistency.

To fix notations, we define Yukawa couplings and the Higgs potential in the SM to be
\begin{eqnarray}
- {\cal L}_{int} & = & \left\{ \bar{e}{\bf F}_L\phi^+l 
                         + \bar{d}{\bf F}_D\phi^+q 
                         + \bar{u}{\bf H}\phi^{+c}q + h.c. \right\} \nonumber \\
            &+&         m^2 \phi^+\phi + {\lambda \over2}(\phi^+\phi)^2 ,
\label{lagr}
\end{eqnarray}  
where three families of fermions are grouped together so 
${\bf F}_L, ~ {\bf F}_D, ~ {\bf H}$
are $3 \times 3$ complex matrices,  and $\phi^c\equiv i\tau_2\phi^*$.
For each coupling constant $x$ in Eq.(\ref{lagr}), we define a corresponding 
$\beta$-function
\begin{equation}
\beta_x = \mu {{d x}\over d \mu} = {1\over16\pi^2} \beta_x^{(1)}
+ {1\over(16\pi^2)^2} \beta_x^{(2)},
\end{equation}
where $\beta_x^{(1)},~ \beta_x^{(2)}$ denote the one-loop and two-loop
contributions, respectively. 
We use dimensional regularization and the modified minimal subtraction scheme 
$(\overline{MS})$ for renormalization.
The expressions of the $\beta_x^{(1)}$'s are quite standard which can be easily reproduced.
The evaluation of $\beta_x^{(2)}$'s will be the object of this
article.                  

Following the conventions of \cite{MV1,MV2,MV3},
we define the following combinations of Yukawa matrices for later convenience
\begin{eqnarray}
 Y_2(S)&=&{\rm Tr} \left[ 3{\bf H^+}{\bf H}+3{\bf F}_D^+ {\bf F}_D + {\bf F}_L^+ {\bf F}_L \right],
\nonumber \\
H(S) &=& {\rm Tr} \left[  3 ({\bf H}^+{\bf H})^2
                              + 3 ({\bf F}_D^+{\bf F}_D)^2 + ({\bf F}_L^+{\bf F}_L)^2 \right],
\nonumber \\ 
 Y_4(S) &=& \left( {17\over20}g_1^2+{9\over4}g_2^2+8g_3^2 \right) {\rm Tr}({\bf H}^+{\bf H}) \nonumber \\
     &+&  \left( {1\over4}g_1^2+{9\over4}g_2^2+8g_3^2 \right) {\rm Tr}({\bf F}_D^+{\bf F}_D ) \nonumber \\
     &+&  {3\over4} \left( g_1^2+g_2^2 \right) {\rm Tr}({\bf F}_L^+{\bf F}_L),
\nonumber \\ 
\chi_4(S)&=&{9\over4}{\rm Tr} \left[ 3({\bf H}^+{\bf H})^2 + 3({\bf F}_D^+{\bf F}_D)^2
         + ({\bf F}_L^+{\bf F}_L)^2 \right. \nonumber \\ 
       &-& \left. {1\over3} \left\{ {\bf H}^+{\bf H}, {\bf F}_D^+{\bf F}_D \right\} \right].
\nonumber 
\end{eqnarray}   
The complex Higgs doublet has to be decomposed into real fields.
Further complication arises since \cite{MV1,MV2,MV3} assumed implicitly
that the fermion fields are real, while usual Weyl fermions are complex. 
Caution should be taken when Yukawa couplings and
gauge representation matrices of fermions are dealt with \cite{luo}.
All issues taken into account, 
$\beta$-functions in the SM can be obtained in a straightforward manner.
The lengthy algebra is greatly simplified 
with the aid of the symbolic software FORM \cite{ver}. 
The $\beta$-functions of the gauge coupling constants are readily reproduced
and conform to those in the literature\cite{MV1}.
Firstly, we present the $\beta$-functions of the Yukawa couplings.
To one loop,
\begin{eqnarray}
  {\bf H}^{-1} \beta_H^{(1)} & = & {3\over2} \left( {\bf H}^+ {\bf H} - {\bf F}_D^+ {\bf F}_D \right)
              + Y_2(S) \nonumber \\
       & - & \left( {17\over20}g_1^2+{9\over4}g_2^2+8g_3^2 \right),  \\
  {\bf F}_D^{-1} \beta_{F_D}^{(1)} &=& {3\over2} \left( {\bf F}_D^+ {\bf F}_D - {\bf H}^+ {\bf H} \right)
              + Y_2(S)  \nonumber \\ 
       & - & \left( {1\over4}g_1^2+{9\over4}g_2^2+8g_3^2 \right),  \\
  {\bf F}_L^{-1} \beta_{F_L}^{(1)} &=& {3\over2}{\bf F}_L^+ {\bf F}_L + Y_2(S)
              - {9\over4} \left( g_1^2+g_2^2 \right);
\end{eqnarray}     
to two loops, 
\begin{eqnarray}
 {\bf H}^{-1} \beta_H^{(2)} &=& {3\over2} ( {\bf H}^+{\bf H} )^2   - {\bf H}^+{\bf H}{\bf F}_D^+{\bf F}_D 
                - {1\over4} {\bf F}_D^+{\bf F}_D{\bf H}^+{\bf H}  \nonumber  \\
      &+& {11\over4} ( {\bf F}_D^+{\bf F}_D )^2 
       + Y_2(S) \left( {5\over4}{\bf F}_D^+{\bf F}_D- {9\over4}{\bf H}^+{\bf H} \right) \nonumber \\ 
      &-& \chi_4(S) + {3\over2}\lambda^2 - 6 \lambda {\bf H}^+ {\bf H}  + {5\over2} Y_4(S) \nonumber \\
      &+& \left( {223\over80} g_1^2 + {135\over16} g_2 ^2 + 16 g_3^2 \right) {\bf H}^+{\bf H} \nonumber \\ 
      &-& \left( {43\over80}g_1^2-{9\over16}g_2^2+16g_3^2 \right) {\bf F}_D^+{\bf F}_D  \nonumber \\
      &+& \left( {9\over200}+{29\over45}n_g \right)g_1^4 
           - {9\over20}g_1^2g_2^2 + {19\over15}g_1^2g_3^2  \nonumber \\
      &-& \left( {35\over4}-n_g \right)g_2^4 
        + 9 g_2^2g_3^2 - \left( {404\over3}-{80\over9}n_g \right) g_3^4,  \label{betaH}\\
 {\bf F}_D^{-1} \beta_{F_D}^{(2)} &=& {3\over2} ({\bf F}_D^+{\bf F}_D)^2 - {\bf F}_D^+{\bf F}_D{\bf H}^+{\bf H} 
                - {1\over4} {\bf H}^+{\bf H}{\bf F}_D^+{\bf F}_D \nonumber  \\ 
      &+&  {11\over4} ({\bf H}^+{\bf H})^2 
      + Y_2(S) \left( {5\over4}{\bf H}^+{\bf H}-{9\over4}{\bf F}_D^+{\bf F}_D \right) \nonumber \\ 
      &-& \chi_4(S) + {3\over2} \lambda^2 - 6\lambda{\bf F}_D^+{\bf F}_D + {5\over2} Y_4(S) \nonumber \\
      &+& \left( {187\over80} g_1^2 + {135\over16} g_2^2 + 16g_3^2 \right) {\bf F}_D^+{\bf F}_D \nonumber \\
      &-& \left( {79\over80}g_1^2-{9\over16}g_2^2+16g_3^2 \right) {\bf H}^+{\bf H}  \nonumber \\
      &-& \left( {29\over200}+{1\over45}n_g \right) g_1^4 
      - {27\over20}g_1^2g_2^2 + {31\over15}g_1^2g_3^2 \nonumber \\
      &-& \left( {35\over4}-n_g \right) g_2^4 
        + 9 g_2^2g_3^2 - \left( {404\over3}-{80\over9}n_g \right) g_3^4, \label{betaFD}\\
 {\bf F}_L^{-1} \beta_{F_L}^{(2)} &=& 
 {3\over2}({\bf F}_L^+{\bf F}_L)^2 - {9\over4}Y_2(S){\bf F}_L^+{\bf F}_L - \chi_4(S) 
                + {3\over2}\lambda^2 \nonumber \\
      &-& 6\lambda{\bf F}_L^+{\bf F}_L 
         + \left( {387\over80}g_1^2+{135\over16}g_2^2 \right) {\bf F}_L^+{\bf F}_L \nonumber \\
      &+& {5\over2} Y_4(S) + \left( {51\over200}+{11\over5}n_g \right)g_1^4 \nonumber \\
      &+& {27\over20}g_1^2g_2^2 - \left( {35\over4}-n_g \right) g_2^4, \label{betaFL}
\end{eqnarray}
where $SU(3) \times SU(2) \times U(1)$ gauge coupling constants $g_3$, $g_2$, and $g_1$  
are normalized based upon $SU(5)$, so the standard electroweak
gauge coupling constants $g$ and $g^{'}$ are related to these by, $g^2=g_2^2$ and
$g^{'2}=3/5g_1^2$. 
The matrices ${\bf F}_L$, ${\bf F}_D$, and ${\bf H}$ do not have to be invertible.
Their inverses should only be understood symbolically and need not be 
introduced in principle.
Properly interpreted, the $\beta$-functions are equal to the Yukawa matrices
themselves multiplied by 
the right-hand side of the corresponding equations.
For $\beta_H^{(2)}$ in \cite{MV2}, 
there was the term $-2 \lambda {\bf H} {\bf F}_D^+{\bf F}_D$,
which is absent in Eq. (\ref{betaH}). 
A close inspection indicates that this term 
arises only from the Feynman diagrams shown in Figs. 1(a) and (b).
However, these two diagrams cancel with each other, 
which can easily be verified by an elementary calculation.
Similarly in Eq. (\ref{betaFD}) for $\beta_{F_D}^{(2)}$, 
the term $-2\lambda {\bf F}_D {\bf H}^+ {\bf H}$ is also absent,
in contrast with \cite{MV2}.
The corresponding Feynman diagrams are shown in Fig. 1(c) and (d) and  
again they cancel with each other.
The leptonic results are included here for completeness.
\begin{figure}[h]  
\begin{center}
\leavevmode
{\epsfxsize=2.50truein \epsfbox{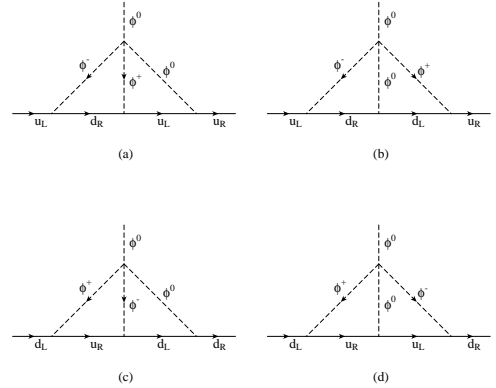}}
\end{center}
\caption{Two-loop diagrams which affect hadronic Yukawa couplings. $(a)$ and $(b)$ 
($(c)$ and $(d)$) cancel with each other,
thus resulting in a null contribution to $\beta_H^{(2)}$ ($\beta_{F_D}^{(2)}$).}
\end{figure}  

The $\beta$-function of the quartic coupling can be obtained in a similar manner.
The calculation is greatly simplified by the fact that there is only one independent
quartic coupling in the SM. 
Calculation for models beyond the SM with numerous quartic couplings 
would be more involved, due to proliferation of combinatorics.
The one-loop contribution to the $\beta$-function of $\lambda$ is
\begin{eqnarray}
  \beta_\lambda^{(1)} & = &12\lambda^2 - \left( {9\over5}g_1^2+9g_2^2 \right) \lambda
                    + \left( {27\over100}g_1^4+{9\over10}g_1^2g_2^2+ {9\over4}g_2^4 \right)
     \nonumber \\
                 &&   + 4Y_2(S) \lambda - 4H(S);
\end{eqnarray}   
to two loops, 
\begin{eqnarray}
\beta_\lambda^{(2)} &=& -78\lambda^3 + \left( 54g_2^2 + {54 \over 5} g_1^2 \right) \lambda^2 
    - \left[ \left( {313\over8} - 10n_g \right) g_2^4 \right. \nonumber \\
   && \left. - {117\over20} g_2^2 g_1^2
       - \left( {687\over200}+ 2n_g \right) g_1^4 \right] \lambda
\nonumber \\
&+&  \left( {497\over8}-8n_g \right) g_2^6 
   - \left( {97\over40}+{8\over5}n_g \right) g_2^4 g_1^2 \nonumber \\
&-& \left( {717\over200}+{8\over5}n_g \right) g_2^2 g_1^4 
   - \left( {531\over1000}+{24\over25}n_g \right) g_1^6 \nonumber \\
&-&  64 g_3^2 {\rm Tr} \left[ ({\bf H}^+{\bf H})^2 + {\bf F}_D^+{\bf F}_D)^2 \right] \nonumber \\
&-& {8\over5} g_1^2 {\rm Tr} \left[ 2({\bf H}^+{\bf H})^2-({\bf F}_D^+{\bf F}_D)^2
                                    +3({\bf F}_L^+{\bf F}_L)^2 \right] \nonumber \\
&-&{3\over2}g_2^4Y_2(S) + g_1^2 \left[  
     \left( {63\over5} g_2^2-{171\over50}g_1^2 \right) {\rm Tr}({\bf H}^+{\bf H}) \right. \nonumber \\
&& \left.   + \left( {27\over5}g_2^2+{9\over10}g_1^2 \right){\rm Tr} ( {\bf F}_D^+{\bf F}_D ) \right. 
\nonumber \\
&&  \left.  + \left( {33\over5}g_2^2-{9\over2}g_1^2 \right) {\rm Tr}({\bf F}_L^+{\bf F}_L)
                     \right]
  + 10\lambda Y_4(S) 
\nonumber \\
&-&  24\lambda^2Y_2(S) - \lambda H(S) 
    - 42\lambda{\rm Tr} \left( {\bf H}^+{\bf H}{\bf F}_D^+{\bf F}_D \right) \nonumber \\
&+&  20 {\rm Tr}\left[ 3({\bf H}^+{\bf H})^3 + 3({\bf F}_D^+{\bf F}_D)^3
                     + ({\bf F}_L^+{\bf F}_L)^3 \right] \nonumber \\
&-& 12{\rm Tr}  \left[ {\bf H}^+{\bf H} 
               \left( {\bf H}^+{\bf H}+{\bf F}_D^+{\bf F}_D \right){\bf F}_D^+{\bf F}_D \right].
\label{betalam}
\end{eqnarray}
Note in Eq. (\ref{betalam}), the coefficient of the term 
$\lambda{\rm Tr} ( {\bf H}^+{\bf H}{\bf F}_D^+{\bf F}_D)$ is $-42$,
instead of $6$ as given by \cite{MV3}.
We note that terms proportional to $\lambda H(S)$ and
$\lambda{\rm Tr} ( {\bf H}^+{\bf H}{\bf F}_D^+{\bf F}_D )$
arise partly from scalar boson propagators.
The relevant Feynman diagrams are shown in Figs. 2(a) and (b).
There are also related two-loop proper scalar quartic vertex diagrams,
which are generically shown in Figs. 2(c) and (d).
It turns out that Fig. 2(c) does not contribute to the $\beta$-function,
so only Figs. 2(a), 2(b), and 2(d) need to be evaluated.
In addition to calculating Feynman diagrams directly, we compare
the coefficient of $\lambda H(S)$ and 
that of $\lambda{\rm Tr} ( {\bf H}^+{\bf H}{\bf F}_D^+{\bf F}_D )$.
By including all specific diagrams and carefully collecting all 
coefficients, we find that the ratio of the term proportional to 
the latter over the term proportional to the former is $42$, instead of $-6$.
This substantiates Eq. (\ref{betalam}).
On the other hand, coefficients of terms $\lambda g_1^2 g_2^2$ and $\lambda g_1^4$
in Eq. (\ref{betalam}) conform to those in \cite{jones},
and in fact Eq. (10) agrees with the corresponding result in \cite{jones}
if only the top-quark Yukawa coupling is retained.
\begin{figure}[h]  
\begin{center}
\leavevmode
{\epsfxsize=2.50truein \epsfbox{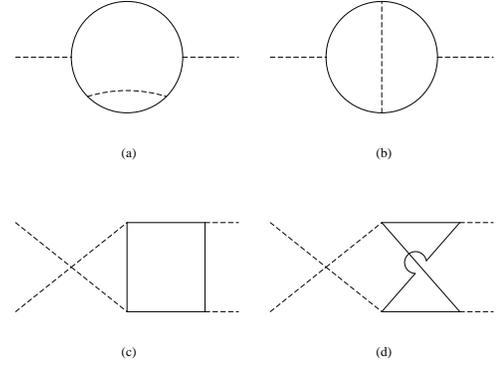}}
\end{center}
\caption{(a) and (b): part of hadronic Yukawa coupling contribution to Higgs boson propagators,
which in turn affect $\beta_\lambda$; (c) and (d): relevant proper scalar quartic vertex diagrams.}
\end{figure}   

The $\beta$-function of $m^2$ can be inferred from the results in \cite{MV3} by introducing
a non-propagating ``dummy" real scalar field $\phi_d$ with no gauge interactions,
and carefully computing the combinatorics associated with the symmetry factor \cite{martin,luo}.
Specifically, the mass term can be re-written as
\begin{eqnarray}
-{\cal L} &=& m^2 \phi^+ \phi = {1\over4!} \lambda_{ddij} \phi_d \phi_d \phi_i \phi_j,
\nonumber
\end{eqnarray}
by decomposing the complex doublet $\phi$ into four real scalars 
$\phi_i, \ (i=1,4) $.
If $\phi_d$ is taken to have no other interactions, then the $\beta$-functions of $m^2$
has the same form as that of the newly introduced quartic coupling $\lambda_{ddij}$.
To one-loop, the $\beta$-function is
\begin{equation}
\beta_{m^2}^{(1)}=m^2 \left[6 {\lambda}+2 Y_2 (S) - {9\over10}g_1^2 -{9\over2}g_2^2 \right],
\end{equation}
and to two-loops
\begin{eqnarray}
\beta_{m^2}^{(2)} & = &     m^2 \left[ -15\lambda^2 -12\lambda Y_2 (S) -{9\over2}H(S) \right. \nonumber \\ 
       &-&    21  {\rm Tr} ( {\bf H}^+{\bf H}{\bf F}_D^+{\bf F}_D)  
             +   \left( {36\over5}g_1^2+36g_2^2 \right)\lambda  + 5Y_4 (S) \nonumber \\ 
&+& \left( n_g +{471\over400} \right) g_1^4 
 \left. +{9\over8}g_1^2 g_2^2 + \left( 5n_g -{385\over16} \right) g_2^4 \right]. \label{betam2}
\end{eqnarray}
Note the terms $n_g g_1^4$ and $n_g g_2^4$  in Eq. (\ref{betam2}), which 
originates from generic Feynman diagrams shown in Fig. 3.
These diagrams are proportional to the fermion Dynkin indices in the pure
fermionic loops, while Dynkin indices are always positive.
So all fermions in the loop contribute additively, thus resulting in
the $n_g$ factor in the expression.
Eq. (\ref{betam2}) is consistent with the newly revised result of \cite{jones}.
\begin{figure}[h]
\begin{center}
\leavevmode
{\epsfxsize=2.50truein \epsfbox{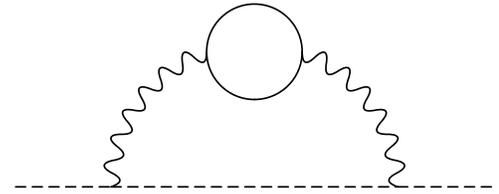}}
\end{center}
\caption{Part of $SU(2) \times U(1)$ contributions to the Higgs boson propagators.}
\end{figure}   

To a good approximation, we can express the SM effective potential of the Higgs field
\begin{equation}
V_{eff}(\phi) = \bar{m}^2(t) Z^2(t) \phi^+ \phi
        + {1\over2} \bar{\lambda}(t) Z^4(t) (\phi^+ \phi)^2,
\end{equation}
in term of the running coupling constants and the running Higgs mass,
\begin{eqnarray}
{d \ln Z(t) \over dt } &=& -  \gamma_\phi(\bar{x}(t))  \nonumber  \\
{ {d\bar{m}^2(t)} \over {dt} } &=& \beta_{m^2}(\bar{x}(t),\bar{m}^2(t)) \nonumber \\
{ {d\bar{x}(t)}   \over {dt} } &=& \beta_{x}(\bar{x}(t))
\end{eqnarray}
where the generic symbol $x$ represents all dimensionless couplings,
including the gauge couplings, the Yukawa couplings and the quartic scalar coupling;
$\gamma_\phi$ is the anomalous dimension of the Higgs field, to be found in \cite{MV1,luo};
$t=\log{\phi/ M}$, where $M$ is an arbitrary mass scale, 
at which the initial values of the coupling constants and the mass are defined.
The vacuum expectation value of the Higgs field is determined by minimization 
of $V_{eff}$.
The physical mass of the Higgs boson is simply the second derivative of $V_{eff}$ 
evaluated at the minimum\cite{higgsmass}.

In summary, we have re-calculated the RGEs in the SM.
A new coefficient is found in the $\beta$-function of the quartic coupling
and a class of gauge invariants are found to be absent in $\beta$-functions of
hadronic Yukawa couplings.
The $\beta$-function of the Higgs mass parameter is also presented in complete form. 
The changes in Yukawa couplings affect the running of the CKM matrix and
the quark masses. 
The changes in $m^2$ and $\lambda$ will change the Higgs potential, 
which in turn affect the triviality and vacuum stability bound on the Higgs mass.
Due to the dominance of one-loop results and relative bigger contributions
from gauge couplings over those from the quartic and Yukawa couplings 
(with the exception of the top quark),
numerical changes are not expected to significant.
For $\lambda$ and Yukawa couplings related the $b$-quark,
the changes are magnified by the factor $m_t^2/M_W^2$, but again
suppressed by the factor $m_b^2/M_W^2$.
All these shifts will be included in a future comprehensive analysis \cite{luo}.
On the other hand, for a heavy fourth family of fermions, the changes would
be sizable.

{\it Note added:} Upon completion of this work, we were informed that part of 
the results presented here were also reached in \cite{haber}.

\centerline{\bf Acknowledgments:}
We are grateful to D. R. T. Jones, S. Martin and M. Vaughn who carefully read the manuscript and
provided independent checks of our results.
We thank J. Vermaseren for helps on FORM.
The work was supported by a Fund for Trans-Century Talents,
CNSF-90103009 and CNSF-10047005.

\end{document}